\documentclass[twocolumn,showpacs,preprintnumbers,amsmath,amssymb]{revtex4}
\usepackage{graphicx}
\usepackage{dcolumn}
\usepackage{bm}
\def\bea {\begin{eqnarray}}
\def\eea {\end{eqnarray}}

\def\be {\begin{equation}}
\def\ee {\end{equation}}

\begin{document}

\title{Transport coefficients of gluonic fluid}

\author{Santosh K Das and Jan-e Alam}

\medskip

\affiliation{Variable Energy Cyclotron Centre, 1/AF, Bidhan Nagar , 
Kolkata - 700064}

\date{\today}
\begin{abstract}
The shear ($\eta$) and bulk ($\zeta$) viscous coefficients have been
evaluated for a gluonic fluid. The elastic, $gg \rightarrow gg$ and the
inelastic, number non-conserving, $gg\rightarrow ggg$ processes 
have been considered as the dominant perturbative processes in evaluating 
the viscous co-efficients to entropy density ($s$) ratios. 
Recently the processes:  $gg \rightarrow ggg$  has been revisited and a
correction to the widely used Gunion-Bertsch (GB) formula has been obtained.  
The $\eta$ and $\zeta$ have been evaluated for gluonic fluid with the formula 
recently derived.  At large $\alpha_s$  the value of $\eta/s$ approaches  
its lower bound, $\sim 1/4\pi$.

\end{abstract}

\maketitle

The nuclear collisions at Relativistic Heavy Ion Collider (RHIC) 
and the Large Hadron Collider (LHC) energies are aimed at creating 
a phase where the properties of the matter is governed 
by the quarks and gluons~\cite{qm08}, such a phase, 
which is mainly composed of light quarks and gluons - is called quark gluon
plasma (QGP). The  weakly interacting picture of the QGP 
stems from the perception of the QCD 
asymptotic freedom at high temperatures and densities.
However, the experimental data  from RHIC, especially 
the measured elliptic flow~\cite{npa2005}
indicate that the matter produced at Au+Au collisions exhibit
properties which are more like strongly interacting liquid than 
a weakly interacting gas.
The shear viscosity or the internal friction of the fluid
symbolizes the ability 
to transfer momentum over a distance of $\sim$mean free path.
Therefore, in a system where the constituents interact strongly 
the transfer of momentum is performed easily 
- resulting in lower values of $\eta$.
Consequently such a system may be characterized by a small value of  
$\eta/s$. The importance of viscosity also lies in the fact that
it damps out the variation in the velocity and make the fluid flow
laminar. A very small viscosity (large Reynold number) may make the
flow turbulent.  

On the other hand  the bulk viscosity exhibit the
exchange of energy between the translational and internal degrees of freedom.  
Although much emphasis has been given to the evaluation
of the shear viscosity for a partonic system recently, 
the bulk viscosity is comparatively less discussed. Probably, because
the bulk viscosity for a structureless point particles vanishes
both for relativistic and non-relativistic limits~\cite{weinberg}. 
However, there are several reasons for which the
bulk viscosity of a system formed in nuclear collisions
at ultra-relativistic energies may be non-zero~\cite{paech}. 
The trace anomaly in QCD will give rise to  non-zero $\zeta$,
which will  indicate the 
deviation of the system from the conformal invariance, because the
$\zeta$  is defined
as the correlation of the trace of the energy momentum tensor through Kubo's 
formula. The $\zeta$  for $SU(3)$ gauge theory has been evaluated 
in lattice QCD and its value is found to be quite large around the
temperature domain for 
partons to hadrons transition ($T_c$)\cite{meyer}. 
The divergence of $\zeta$ may be treated as a signal of critical point
as it diverges near this point~\cite{kharzeev}. However, this point
has been challenged in~\cite{GDM}.
This indicate that both $\eta$ and $\zeta$ can be used effectively
to characterize QGP.  Therefore,
in the present work we would like to  estimate both the ratios, $\eta/s$
and $\zeta/s$   
for a gluonic fluid by taking into account the perturbative 
QCD (pQCD) elastic process, $gg\rightarrow gg$ and the 
inelastic, number non-conserving process, $gg\rightarrow ggg$~\cite{FAB}.
While evaluating the transport coefficients we will use 
the newly obtained matrix element for $gg\rightarrow ggg$
process~\cite{DA}.

\begin{figure}[h]
\begin{center}
\includegraphics[scale=0.43]{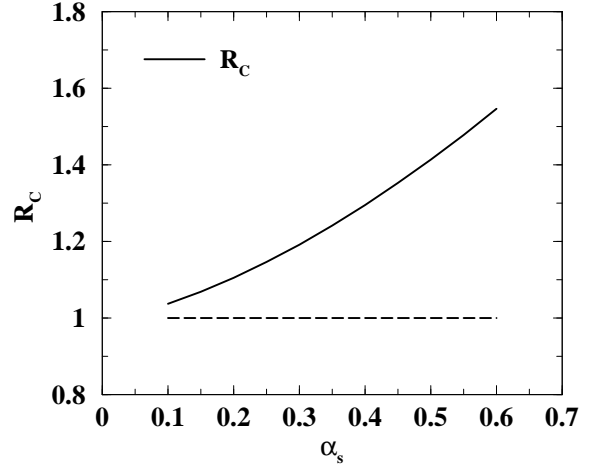}
\caption{The variation of the quantity, 
$ R_c$ (Eq.~{\protect{\ref{eq1}}}) with coupling constant.}
\label{fig1}
\end{center}
\end{figure}

The transport coefficients for QCD matter has been evaluated 
in~\cite{hosoya,sgavin,PDMG}. The calculation of the viscous coefficients
within the ambit of diagrammatic approach of 
quantum field theory, along with its limitation 
has been discussed in ~\cite{jeon}.
Recently pQCD approaches~\cite{AMY,XG,XG1,CDOW,AMYzeta} have been used to calculate 
$\eta/s$.  Evaluation of $\eta/s$ for 
a gluonic plasma by Xu and Greiner(XG) indicates that the contribution 
from gg $ \rightarrow$ ggg is 7 times as 
large as that from   gg $ \rightarrow$ gg. This 
brings the value of $\eta/s $ down to the AdS/CFT bound 
$\sim 1/4\pi$~\cite{KSS},
when the strong coupling constant, $ \alpha_s=0.6 $. 
The GB  formula~\cite{GB}(see also~\cite{wong}) 
for the $gg\rightarrow ggg$ matrix element 
squared is used in~\cite{XG}. However,   we have shown recently 
that at the lower temperature domain 
the GB formula receives a significant correction.
The ratio of matrix element squared with~\cite{DA}(henceforth will be denoted by
the subscript DA) and without~\cite{GB}(henceforth denoted by the subscript GB)
the correction term is given by: 
\begin{equation}
R_c=\frac{\left| M_{gg \rightarrow ggg} \right|^2_{\mathrm DA}}{\left| M_{gg \rightarrow ggg} \right|^2_{\mathrm GB}×}=
1+\frac{(q_\perp^2+m_D^2)^2}{s^2×}
\label{eq1}
\end{equation}

where
\begin{eqnarray}
\left| M_{gg \rightarrow ggg} \right|^2_{\mathrm DA}=\left(\frac{4g^4N_c^2}{N_c^2-1×}
\frac{s^2}{(q_\perp^2+m_D^2)^2×}\right) \nonumber \\
\left(\frac{4g^2N_cq_\perp^2}{k_\perp^2\lbrack(k_\perp-q_\perp)^2+m_D^2\rbrack} \right) \nonumber \\
+\frac{16g^6N_c^3}{N_c^2-1×}\frac{q_\perp^2}{k_\perp^2\lbrack(k_\perp-q_\perp)^2+m_D^2\rbrack}
\label{eq2}
\end{eqnarray}
and
\begin{eqnarray}
\left| M_{gg \rightarrow ggg} \right|^2_{\mathrm GB}=\left(\frac{4g^4N_c^2}{N_c^2-1×}
\frac{s^2}{(q_\perp^2+m_D^2)^2×}\right) \nonumber \\
\left(\frac{4g^2N_cq_\perp^2}{k_\perp^2\lbrack(k_\perp-q_\perp)^2+m_D^2\rbrack} \right) 
\label{eq3}
\end{eqnarray}
where $g$ is the colour charge, $N_c$ is the number of the colour,
$m_D\sim gT$ is the thermal (Debye) 
mass of the gluon, $k_\perp$ is the transverse momentum of the 
emitted gluon and $q_\perp$ is the transverse momentum of the 
exchanged gluon.  Following our previous work~\cite{DA} we depict the 
magnitude of the correction, $R_c$ in Fig.~\ref{fig1}.
The values of $q_\perp$ and $s$ are same as the values 
taken in Ref.~\cite{DA}.
It is observed that for large values of $\alpha_s$ the corrections to the 
GB matrix element is significant. Therefore, it is expected that the
values of energy loss, $\eta/s$ and $\zeta/s$ will also be affected by the
correction term in the lower temperature (higher coupling) domain.

\begin{figure}[h]
\begin{center}
\includegraphics[scale=0.43]{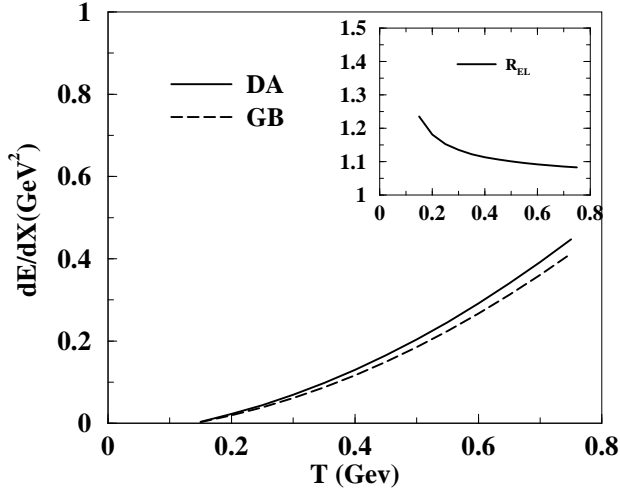}
\caption{The variation of energy loss with temperature for the
process: $gg\rightarrow ggg$. Inset:
The Variation of $ R_{EL}$ (Eq.~\ref{eq5}) with temperature. 
}
\label{fig2}
\end{center}
\end{figure}

Before discussing the bulk and  shear viscosities 
we estimate the effects of the correction term to the 
radiative energy loss mechanism  of partons propagating through QGP
which is measured experimentally thorough the
nuclear suppression factor~\cite{jetq}
in heavy ion 
collision. To evaluate the radiative energy loss we start 
with the soft gluon distribution, which can be written as~\cite{DA}
\begin{eqnarray}
\frac{dn_g}{d\eta dk_\perp^2×}=\frac{C_A\alpha_s}{\pi^2} 
\left(\frac{q_\perp^2}{k_\perp^2\lbrack(k_\perp-q_\perp)^2+m_D^2\rbrack×} \right) + \nonumber \\
\frac{C_A\alpha_s}{\pi^2}
\left(\frac{q_\perp^2(q_\perp^2+
m_D^2)^2}{s^2k_\perp^2\lbrack(k_\perp-q_\perp)^2+m_D^2\rbrack×} \right)
\label{eq4}
\end{eqnarray}
where $k=(k_0,k_\perp,k_3)$ is the four momenta of 
the emitted gluon,  $q=(q_0,q_\perp,q_3)$ is the four momenta of 
the exchanged gluon and
$C_A=3$ is the Casimir invariant of the $SU(3)$ adjoint representation.
The $m_D$ in Eq.~\ref{eq3} is the Debye mass required to shield the infra-red
divergence. 
We use the above spectrum of the soft gluon to evaluate the
radiative energy loss, $dE/dx$   of gluons. 
The minimum value of the momentum of the exchanged gluon in the 
process: $gg\rightarrow ggg$ sets time (length) scale for the formation 
time of the emitted gluon. If the formation time is comparable to 
or larger than the mean free time (path) then the scattering of the gluons from
the successive scatterers in the medium can not be treated as independent
and the scattering amplitudes between two adjacent interaction may
interfere destructively leading to the  the suppression of 
the emission process - called LPM suppression.
The LPM suppression  
has been taken into account by including a formation time restriction
on the phase space of the emitted gluon in which the formation time, $\tau_F$
must be smaller than the interaction time, $\tau=\Gamma^{-1}$,
$\Gamma^{-1}$ being the interaction rate. The radiative
energy loss of heavy quark can be given by~\cite{DAM2}: 
\be
-\frac{dE}{dx}\mid_{\mathrm rad}=\Gamma\epsilon=\tau^{-1}.\epsilon
\label{dedx}
\ee
where $\epsilon$, the average energy per collision is given by
\be
\epsilon = \textless n_gk_0 \textgreater=\int d\eta d^2k_\perp
\frac{dn_g}{d\eta d^2k_\perp×} k_0 \Theta(\tau-\tau_F)
\ee
where $\tau_F={\mathrm cosh}\eta/k_\perp$.
The $dE/dx$ is evaluated with
temperature dependent $\alpha_s$.
The variation of $\alpha_s$ (and hence $m_D^2\sim \alpha_s(T)T^2$)
with $T$ has been taken from Ref.~\cite{zantow}.  
In Fig.~\ref{fig2} the variation of radiative energy loss  
with $T$ has been depicted for the process $gg\rightarrow ggg$. 
The solid line (dotted line) represents the energy loss when DA (GB) gluon 
multiplicity distributions are used. To emphasize the
importance of the corrections to GB formula we display the
ratio,
\begin{eqnarray}
R_{EL}=\frac{\mathrm{DA}_{EL}}{\mathrm{GB}_{EL}×}
\label{eq5}
\end{eqnarray}
in the inset of Fig.~\ref{fig2}.
It is observed that the correction to the gluon spectrum, which leads 
to the energy loss is appreciable for lower temperature domain.  This may
affect the suppression of high $p_T$ partons in QGP and the elliptic flow 
of the matter formed at RHIC~\cite{npa2005} and LHC~\cite{alicev2} energies. 

\begin{figure}[h]
\begin{center}
\includegraphics[scale=0.43]{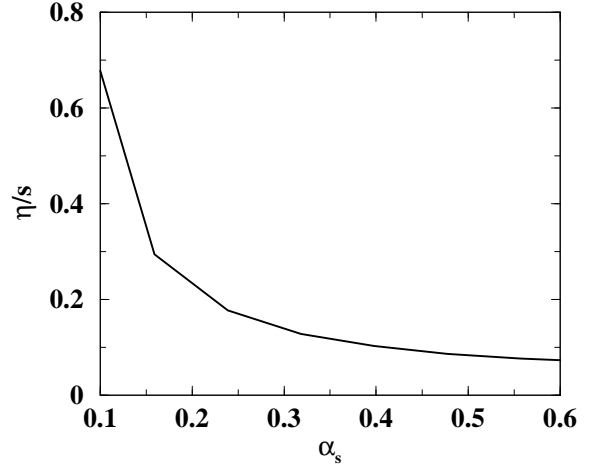}
\caption{Variation of $\eta/s$  with strong coupling constant. 
}
\label{fig3}
\end{center}
\end{figure}

We  calculate the quantity, $\eta/s$ for a gluonic 
system  for the pQCD processes: $gg\rightarrow gg$,  $gg\rightarrow ggg$
and $ggg\rightarrow gg$.
The $\eta$ is evaluated using the procedure outlined in~\cite{CDOW}
(see also ~\cite{CDDW} for details):
\begin{eqnarray}
\eta &=&\frac{N_{g}^{2}\beta }{80}\int \prod_{i=1}^{4}\frac{d^{3}\mathbf{k}%
_{i}}{(2\pi )^{3}2E_{i}}|M_{gg\rightarrow gg}|^{2}\notag\\
&&\times (2\pi )^{4}\delta^{4}(k_{1}+k_{2}-k_{3}-k_{4})\notag\\
&&\times (1+f_{1}^{0}) (1+f_{2}^{0})f_{3}^{0}f_{4}^{0}\notag \\
&&\times \lbrack B_{ij}(k_{4})+B_{ij}(k_{3})-B_{ij}(k_{2})-B_{ij}(k_{1})]^{2}\notag \\
&&+\frac{N_{g}^{2}\beta }{120}\int \prod_{i=1}^{5}\frac{d^{3}\mathbf{k}_{i}}{%
(2\pi )^{3}2E_{i}}|M_{gg\rightarrow ggg}|^{2}\notag\\
&&\times (2\pi )^{4}\delta^{4}(k_{1}+k_{2}-k_{3}-k_{4}-k_{5})\notag\\
&&\times (1+f_{1}^{0})(1+f_{2}^{0})f_{3}^{0}f_{4}^{0}f_{5}^{0}\notag\\
&&\times \lbrack B_{ij}(k_{5})+B_{ij}(k_{4})+B_{ij}(k_{3})\notag\\
&&-B_{ij}(k_{2})-B_{ij}(k_{1})]^{2}\notag\\
\label{eta}
\end{eqnarray}
where $B_{ij}(k)\equiv B(k)(\hat{p}^{i}\hat{k}^{j}-\frac{1}{3}\delta _{ij})$,
is defined through the infinitesimal deviation from the equilibrium value
of the gluon phase space density.
For the process:  $gg\rightarrow ggg$ we use the matrix element obtained
recently in~\cite{DA}.
The variation of $\eta/s$ with $s=16\times 2\pi^2 T^3/45$ 
as a function of $\alpha_s$  
is depicted in Fig.~\ref{fig3}. The $\eta/s$ is quite large
at low $\alpha_s$ because for weakly interacting system the momentum
transfer between the constituents become strenuous which give rise
to large $\eta$. However, with the increase in the
coupling strength the momentum transfer gets easier as a result
the shear viscosity reduces. The results 
indicate that for large $\alpha_s$ the quantity, 
$\eta/s$ approaches the AdS/CFT limit. However, in such a 
scenario the non-perturbative effects  may become important. 
This can be verified by performing a lattice QCD based calculations 
(which include the non-perturbative effects) for
pure SU(3) gauge theory. In fact, such  calculation of $\eta/s$ has been 
done in ~\cite{meyer1} and it is found that the value is
close to AdS/CFT bound. 

As mentioned before the bulk viscosity, 
which is connected with the trace of the energy momentum tensor
through Kubo's formula, will be non-zero for a system where
the conformal symmetry is broken. 
Lattice QCD calculations indicates non-zero $\zeta$ 
for a gluonic plasma~\cite{meyer} due to purely quantum effects (trace anomaly)
(see also ~\cite{dg,bazavov} for QGP and ~\cite{fernandez} for
pions) for temperatures around $T_c$.
Physically, the bulk viscosity appears in the processes 
which are accompanied by 
a change in the volume(i.e. in density) of the fluid. In compression 
or expansion, as in any rapid change of state, the fluid ceases to 
be in thermodynamic equilibrium, and internal processes are set up in 
it which tend to restore the equilibrium. But the processes which 
drives the system toward equilibrium are irreversible associated with the increase in  entropy 
and therefore involve energy dissipation. Hence, if the relaxation 
time of these processes is long, a considerable dissipation of 
energy occurs when the fluid is compressed or expanded and this 
dissipation must be determined by the bulk viscosity~\cite{landau}.
We evaluate the bulk viscous coefficient with the following formula
(see Refs.~\cite{KTV,SR,KTV1,KTV2,PC}):
\be
\zeta=\frac{\mathrm deg}{T}
\int\frac{p^2dp}{2\pi^2}\frac{1}{\Gamma(p)}f_p(1+f_p)
\left[\delta c_s^2E\right]^2
\label{eq7}
\ee
where ${\mathrm deg}$ is the statistical degeneracy for the gluons, 
$f_p$ is the Bose-Einstein distribution for the gluons, 
$\Gamma$ is the interaction rate, evaluated using the 
techniques similar to the one outlined in~\cite{CDOW} with the matrix 
elements $M_{gg\rightarrow gg}$ and $M_{gg\rightarrow ggg}$ for the 
processes $gg\rightarrow gg$ and $gg\rightarrow ggg$ respectively,
$c_{s}^2$ is the velocity of sound and $\delta c_s^2=(1/3-c_s^2)$.
The value of the velocity of sound, $c_s$ 
for a massless system in equilibrium is $1/\sqrt{3}$,
therefore, the results indicate that the bulk viscosity vanishes 
for a massless system in equilibrium. 
As $\delta c_s^2$  is a measure of the deviation from conformal symmetry 
(for massless system) the $\zeta$ increases with $\delta c_s^2$.

\begin{figure}[h]
\begin{center}
\includegraphics[scale=0.43]{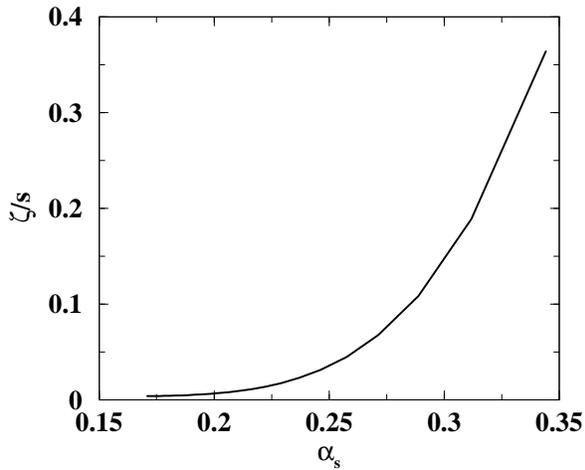}
\caption{The variation of $\zeta/s$ with the strong coupling constant. 
The temperature dependence gluon mass is included here.
}
\label{fig4}
\end{center}
\end{figure}

We evaluate the bulk viscosity with 
the temperature dependent $c_s^2$~\cite{lqcdcs}.
In Fig.~\ref{fig4}, the ratio, $\zeta/s$ is depicted as a function of
strong coupling.  The results shown here contain temperature dependence
thermal gluonic mass. 
The most striking observation one can make here is the
completely different kind of variation of $\eta/s$ and $\zeta/s$
with $\alpha_s$. While $\zeta/s$ increases with $\alpha_s$~\cite{AMYzeta},
the $\eta/s$ reduces with it~\cite{AMY}. At small $\alpha_s$
the bulk viscosity is negligibly small.

We have evaluated the shear and bulk viscosities for a gluonic 
system including the pQCD processes: $gg\rightarrow gg$ and
$gg\rightarrow ggg$. The matrix element for the later processes
is taken from ~\cite{DA}. We find that 
the value of $\eta/s$ approaches the AdS/CFT lower bound 
for large $\alpha_s$.
The value of $\eta/s=0.12$ obtained here at $\alpha_s=0.3$ is within the limit
extracted from the analysis of elliptic flow of matter formed 
in nuclear collisions at RHIC energy~\cite{rhicexpt}. 
The value of $\zeta/s\sim 0.15$  for $\alpha_s\sim 0.3$, which is 
comparable to $\eta/s$ at the same value of $\alpha_s$.

For $\alpha_s=0.47$ where the corrections to GB formula 
is large, we get the AdS/CFT lower bound $\eta/s=1/4\pi$. 
Now if we want to understand what is the magnitude of $\eta/s$
realized in the partonic  matter expected to be formed 
in heavy ion collisions at RHIC and LHC
energies, then one possible way is to first estimate the 
values of temperatures attained in these collisions and
then needs to know what is the value of the $\alpha_s$ 
corresponding to these temperatures.  The variation of strong
coupling with temperature may be taken, for example, from~\cite{zantow},
for this purpose.
The typical values of temperatures which can be achieved in heavy ion
collisions at RHIC and LHC energies are $\sim 300$ MeV and $700$ MeV
respectively.  
The magnitudes of $\alpha_s$ at these temperatures are of the 
order of 0.23 and 0.17 respectively. From the variation of
$\eta/s$ with $\alpha_s$ (Fig.~\ref{fig3}) we conclude that 
the typical values of $\eta/s$ which may realized at RHIC and LHC 
collisions are 0.177 and 0.29 respectively, 
which is above the AdS/CFT bound but 
close to the value obtained from the analysis of experimental data at RHIC
~\cite{rhicexpt}.

{\bf Acknowledgment:} 
We are grateful to   Qun Wang and  Hui Dong for useful discussions
and helping in getting the results on shear viscosity.  
We are also thankful to Andrey Khvorostukhin and Purnendu Chakrabarty
for useful comments. This work is supported by DAE-BRNS project Sanction No.  
2005/21/5-BRNS/2455.

\end{document}